\documentclass[aps,showpacs]{revtex4}

\usepackage{graphicx}
\usepackage{array}
\usepackage{bm}
\usepackage[english]{babel}
\usepackage{dcolumn}
\usepackage{amssymb}
\usepackage{amsmath}

\begin{document}

\title{An upper bound for asymmetrical spinless Salpeter equations}

\author{Claude \surname{Semay}}
\thanks{F.R.S.-FNRS Senior Research Associate}
\email[E-mail: ]{claude.semay@umons.ac.be}
\affiliation{Service de Physique Nucl\'{e}aire et Subnucl\'{e}aire,
Universit\'{e} de Mons,
Acad\'{e}mie universitaire Wallonie-Bruxelles,
Place du Parc 20, 7000 Mons, Belgium}

\date{\today}

\begin{abstract}
A generic upper bound is obtained for the spinless Salpeter equation with two different masses. Analytical results are presented for systems relevant for hadronic physics: Coulomb and linear potentials when a mass is vanishing. 
\end{abstract}


\maketitle

\section{Introduction}
\label{intro}

The spinless Salpeter equation (SSE), whose general form is given by ($\hbar = c = 1$)
\begin{equation}
\label{SSE12}
\left( \sqrt{\bm p^2 +m_1^2} + \sqrt{\bm p^2 +m_2^2} + V(\bm r) \right) |\phi\rangle = M |\phi\rangle,
\end{equation}
where $M$ is the mass of the system, is the simplest relativistic eigenvalue equation. It is sometimes denoted semirelativistic since it is not a covariant formulation. This equation can be considered as a Schr\"odinger equation with its nonrelativistic kinetic part replaced by a relativistic counterpart. More rigorously, it is obtained from the covariant Bethe-Salpeter equation \cite{salp51} with the following approximations: Elimination of any dependences on timelike variables and neglect of  particle spin degrees of freedom as well as negative energy solutions \cite{grei94}. It has also been shown that a Nambu-Goto string meson with a short-range Coulomb interaction is described by a SSE with a time component vector funnel potential, for fixed angular momentum and large radial excitation \cite{alle03}.

Numerous techniques have been developed to solve numerically this equation \cite{nick84,godf85,fulc94,brau98,sema01}. Though a quite high precision can be reached, it is always interesting to obtain bounds on eigenvalues. First, they can be used as verifications of the computation. Second, they are generally a simple and fast way to obtain information about the spectra. Sometimes, bounds yield analytical results which can give precious details about the solutions of the SSE. With some rare exceptions \cite{brau98b,silv09}, most of the analytical results have been obtained for the symmetrical version of the SSE \cite{dura83,hall01,hall01b,hall02,hall03,hall05,brau05,brau05b,brau05c,hall07}
\begin{equation}
\label{SSEsig}
\left( \sigma \sqrt{\bm p^2 +m^2}+ V(\bm r) \right) |\phi\rangle = M |\phi\rangle.
\end{equation}
The one-body (two-body) case is treated with $\sigma=1$ (2). Note that an arbitrary positive value of $\sigma$ can be considered to study duality relations between different many-body systems \cite{silv11}. 

In this work, we will use the auxiliary field method (AFM) \cite{silv09,silv11,silv08,sema11} to obtain a generic upper bound on the eigenvalues of the general SSE. The computation are presented in section~\ref{UB} after a brief description of the AFM. In section~\ref{AR}, some analytical results obtained with this bound are given and compared with numerical solutions. Concluding remarks are given in section~\ref{conclu}.

\section{Upper bound}
\label{UB}

At the origin, the AFM was introduced to get rid of the square root kinetic operator in calculations for semirelativistic eigenvalue equations \cite{morg99,kala00}. More recently, it has been generalized to treat a greater variety of problems \cite{silv09,silv11,silv08,sema11}. As it is shown in \cite{buis09}, the AFM has strong connections with the envelope theory \cite{hall01,hall01b,hall02,hall89}. Nevertheless, both methods have been introduced from completely different starting points. In particular, the AFM introduces the notion of auxiliary fields (also called einbein fields) which is explained below. 

We will consider here a potential $V(r)$ depending only on the radial distance. The idea is to replace the Hamiltonian in (\ref{SSE12}) by the following one
\begin{equation}
\label{Htilde}
\tilde H = \frac{\nu_1^2+m_1^2}{2 \nu_1} + \frac{\nu_2^2+m_2^2}{2 \nu_2} +\frac{\bm p^2}{2 \mu} + \rho\,P(r) +  V(I(\rho)) -\rho\,P(I(\rho)),
\end{equation}
with three auxiliary fields $\nu_1$, $\nu_2$ and $\rho$, and  
\begin{equation}
\label{Htilde2}
\mu=\frac{\nu_1\,\nu_2}{\nu_1+\nu_2}, \quad I(x)=K^{-1}(x), \quad K(x)=\frac{V'(x)}{P'(x)}.
\end{equation}
The kinematics is nonrelativistic, $P(r)$ is the auxiliary potential and the prime denotes the derivative. If we put in $\tilde H$ the field values which extremize $\tilde H$,
\begin{equation}
\label{extrema}
\nu_i =\sqrt{\bm p^2 +m_i^2} \quad  \textrm{and} \quad \rho = K(r),
\end{equation}
then $\tilde H$ reduces to the genuine Hamiltonian in (\ref{SSE12}). The approximation lies in the fact that these auxiliary fields are no longer considered as operators but as constants. In this case, an eigenvalue of $\tilde H$ is given by 
\begin{equation}
\label{EAFM}
M(\nu_1,\nu_2,\rho) = \frac{\nu_1^2+m_1^2}{2 \nu_1} + \frac{\nu_2^2+m_2^2}{2 \nu_2}
 +V(I(\rho)) - \rho\,P(I(\rho)) + \epsilon(\mu,\rho),
\end{equation}
where $\epsilon(\mu,\rho)$ is an eigenvalue of the nonrelativistic Hamiltonian
\begin{equation}
\label{HNR}
H_{\textrm{\scriptsize{NR}}} =  \frac{\bm p^2}{2 \mu}+ \rho\,P(r).
\end{equation}
The AFM approximation for an eigenvalue of the genuine Hamiltonian is given by $M(\nu_1,\nu_2,\rho)$ for which 
\begin{equation}
\label{condextrema}
\frac{\partial M(\nu_1,\nu_2,\rho)}{\partial \nu_i} = \frac{\partial M(\nu_1,\nu_2,\rho)}{\partial \rho} = 0.
\end{equation}

If $V(r) \propto P(r)$, the treatment of the potential is trivial and the AFM reduces to the replacement of the semirelativistic kinetic energy by the best possible nonrelativistic counterpart. An AFM eigenvalue is then an upper bound \cite{silv09}. In the general case, a function $g(x)$ can be defined by $V(x) = g(P(x))$. It can then be shown that an AFM eigenvalue is an upper bound on the exact energy if $g(x)$ is a concave function ($g''(x) < 0$). This property has been demonstrated in the framework of the envelope theory \cite{hall01,hall01b,hall02}, but can be applied as well to the AFM \cite{buis09}. 

Interesting results can be obtained if the auxiliary potential is a power law,
\begin{equation}
\label{Pr}
P(r) = \textrm{sgn}(p)\, r^p \quad \textrm{with} \quad p > -2,
\end{equation}
whose sign is chosen in order that $\rho$ is always a positive quantity. In this case, a very convenient parameterization of $\epsilon(\mu,\rho)$ is given by \cite{silv08,hall89}
\begin{equation}
\label{epsPr}
\epsilon(\mu,\rho) = \frac{p+2}{2 p}\left( |p|\rho \right)^{2/(p+2)}\left(\frac{Q^2}{\mu}\right)^{p/(p+2)}. 
\end{equation}
$Q$ is a global quantum number which is exactly known in the following cases:
\begin{itemize}
\item $p=2$, $Q=Q_{2}=2\, n+l+3/2$;
\item $p=1$, $Q=Q_{1}=2 (-\alpha_n/3)^{2/3}$, for $l=0$, where $\alpha_n$ is the $(n + 1)$th zero of the Airy function Ai;
\item $p=-1$, $Q=Q_{-1}=n+l+1$.
\end{itemize}
In the general case, accurate values of $Q$ can be easily numerically computed. Simple and good analytical approximations can be found in \cite{silv08}.

After some algebra, constraints (\ref{condextrema}), with $P(r)$ given by (\ref{Pr}), reduce to 
\begin{eqnarray}
\label{cond1}
&&\nu_i^2 - m_i^2 = \left( |p|\, K(r_0)\, \mu \right)^{2/(p+2)} Q^{2 p/(p+2)}, \\
\label{cond2}
&&r_0^p = \left( |p|\, K(r_0)\, \mu \right)^{-p/(p+2)} Q^{2 p/(p+2)},
\end{eqnarray}
where $r_0$ is defined by $r_0= I(\rho)$. From (\ref{cond1}) and (\ref{cond2}), we can deduce that
\begin{equation}
\label{nui}
\nu_1^2 - m_1^2 = \nu_2^2 - m_2^2 = \left( \frac{Q}{r_0} \right)^2.
\end{equation}
It seems then quite natural to define $p_0 = Q/r_0$ in such a way that $\nu_1=\sqrt{p_0^2+m_1^2}$ and $\nu_2=\sqrt{p_0^2+m_2^2}$. Taking into account that $K(r_0) = V'(r_0)/(|p|r_0^{p-1})$, (\ref{cond1}) and (\ref{cond2}) can be written as
\begin{equation}
\label{condfin}
Q^2 \left( \frac{1}{\nu_1}+\frac{1}{\nu_2} \right) = r_0^3 V'(r_0).
\end{equation}
After some algebra, the approximate eigenvalue (\ref{EAFM}) with the parameterization (\ref{epsPr}) can be written
\begin{equation}
\label{condfin2}
M = \nu_1+\nu_2+V(r_0).
\end{equation}

Finally, the AFM solution is given by the following set of equations
\begin{eqnarray}
\label{AFM1}
&&M = \sqrt{p_0^2+m_1^2}+\sqrt{p_0^2+m_2^2}+V(r_0), \\
\label{AFM2}
&&p_0 = \frac{Q}{r_0}, \\
\label{AFM3}
&&\frac{p_0^2}{\sqrt{p_0^2+m_1^2}} + \frac{p_0^2}{\sqrt{p_0^2+m_2^2}} = r_0 V'(r_0).
\end{eqnarray}
The parameter $r_0$ can then be interpreted as a mean distance between the particles and $p_0$ as a mean momentum per particle. If $g(x)$, with $V(x) = g(\textrm{sgn}(p)\, x^p)$, is a concave function or if $V(x) \propto \textrm{sgn}(p)\, x^p$, then $M$ is an upper bound on a true eigenvalue. It is remarkable that the only trace of the auxiliary potential is contained in the value of $Q$. With the notation, $T(x)=\sqrt{x^2+m_1^2}+\sqrt{x^2+m_2^2}$, the system (\ref{AFM1})-(\ref{AFM3}) can be recast under a form similar to the one of the system (3.2)-(3.4) in \cite{sema11}. This shows that (\ref{AFM3}) is the translation into the AFM variables 
of the generalized virial theorem \cite{virial}.

\begin{figure}[ht]
\begin{center}
\includegraphics*[width=6cm]{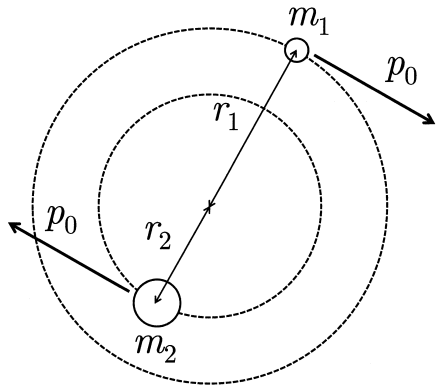}
\caption{Classical circular motion of the two relativistic particles. \label{fig:class}}
\end{center}
\end{figure}

A semiclassical interpretation of the system (\ref{AFM1})-(\ref{AFM3}) is also possible. Let us assume a classical circular motion for the two particles, as illustrated on fig.~\ref{fig:class}. In this case, the force $F_i$ acting on particle $i$ is given by
\begin{equation}
\label{Fi}
F_i = m_i\, \gamma(v_i)\, \frac{d v_i}{dt} \quad \textrm{with} \quad \frac{d v_i}{dt} = \frac{v_i^2}{r_i}.
\end{equation}
Taking into account that both particles are characterized by the same momentum $p_0$, This equation becomes
\begin{equation}
\label{Fi2}
F_i =\frac{p_0^2}{\sqrt{p_0^2+m_i^2}}\frac{1}{r_i}.
\end{equation}
The rigid rotation constraint, $v_1/r_1 = v_2/r_2$, implies that 
\begin{equation}
\label{r0r1r2}
r_0 = r_1 + r_2 = r_i \frac{\sqrt{p_0^2+m_1^2}+\sqrt{p_0^2+m_2^2}}{\sqrt{p_0^2+m_j^2}},
\end{equation}
with $i\ne j$. If the force acting on $i$ comes from the potential $V(r)$ generated by $j$, then $F_1=F_2=V'(r_0)$. (\ref{Fi2}) and (\ref{r0r1r2}) can be recast onto the form (\ref{AFM3}), and it is obvious than (\ref{AFM1}) gives the mass of the system. The total orbital angular momentum is $r_1\, p_0+r_2\, p_0=r_0\, p_0$. A semiclassical quantification gives thus $r_0\, p_0 = L+1/2$, and we obtain a system very similar to (\ref{AFM1})-(\ref{AFM3}). Nevertheless, the AFM produces more general results: Equations are obtained not only for a circular motion; The quantum number $Q$ is unambiguously determined by the choice of the power law auxiliary potential; A good choice of this potential can allow to obtain an upper bound; An eigenstate of $H_{\textrm{\scriptsize{NR}}}$ (\ref{HNR}), with $\nu_i=\sqrt{p_0^2+m_i^2}$ and $\rho=K(r_0)$, is an approximation of the genuine eigenstate \cite{eigen}.

\section{Analytical results}
\label{AR}

It is easy to obtain numerical upper bounds of the SSE with the system (\ref{AFM1})-(\ref{AFM3}). One can wonder if it is possible to obtain analytical solutions? Due to the presence of the square roots, it is only conceivable with very special conditions. If one mass is vanishing ($m_1=0$ and $m_2=m$) and if the potential is a sum of power-law interactions $V(r) = \sum_i \textrm{sgn}(\lambda_i)\,\alpha_i\, r^{\lambda_i}$ with $\alpha_i \ge 0$, (\ref{AFM2})-(\ref{AFM3}) reduce to
\begin{equation}
\label{AFM3fun}
Q + \frac{Q^2}{\sqrt{Q^2+m^2 r_0^2}} = \sum_i |\lambda_i|\,\alpha_i\, r_0^{\lambda_i+1}.
\end{equation}
This transcendental equation can only be easily solved for $V(r)=-a/r+b\, r$, in such a way that only $r_0^2$ appears in (\ref{AFM3fun}). This corresponds to the funnel potential often used in hadronic physics: The sort range part is a strong Coulomb interaction stemming from the one-gluon exchange mechanism while the long range part is a linear confinement \cite{alle03,godf85,fulc94,morg99,kala00}. In this case, (\ref{AFM3fun}) is a cubic equation in $r_0^2$ whose solution is a very complicated expression, unusable in practice. So we will only consider the two parts of the funnel potential separately.

\subsection{Coulomb potential}
\label{cp}

\begin{figure}[ht]
\begin{center}
\includegraphics*[width=0.45\textwidth]{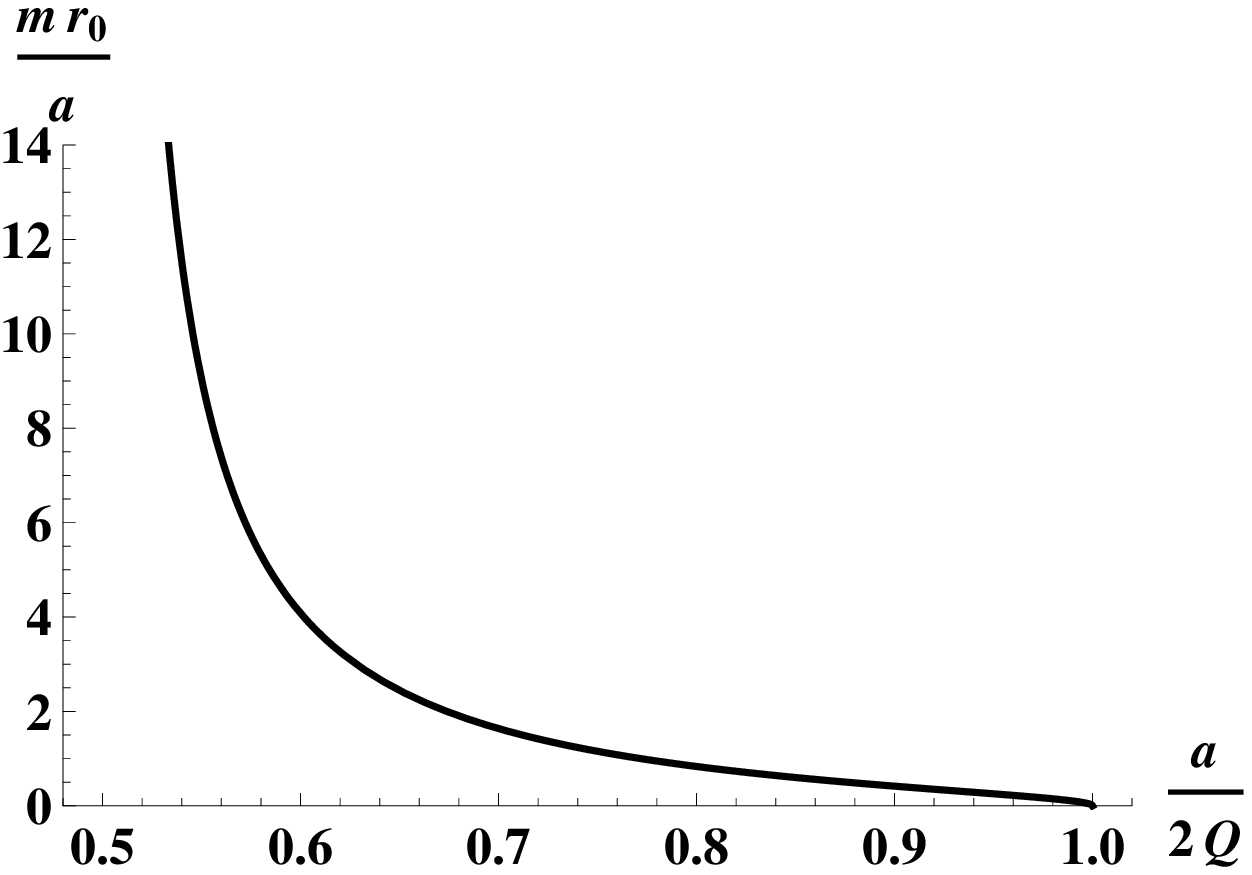}
\qquad
\includegraphics*[width=0.45\textwidth]{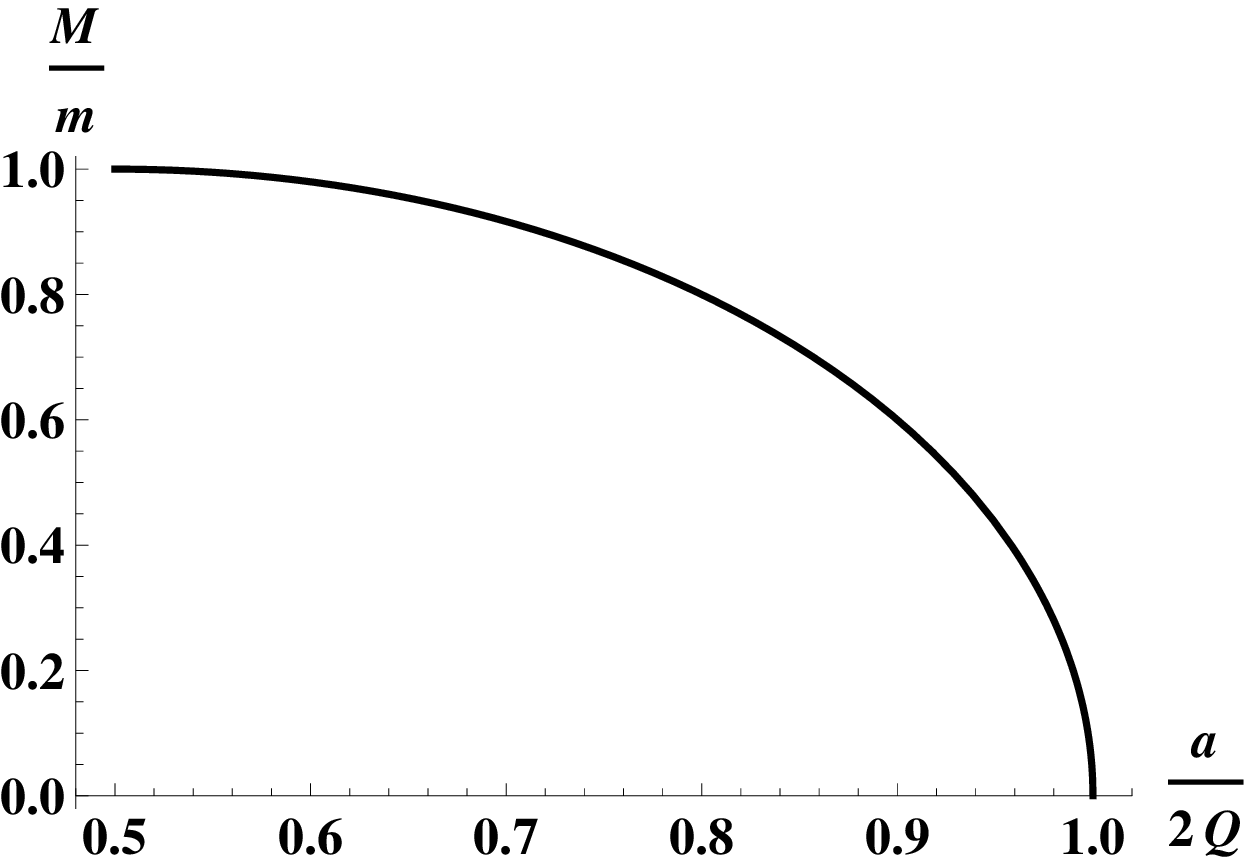}
\caption{AFM solution with a global quantum number $Q$ for a particle of mass $m$ interacting with a massless particle via a Coulomb potential $-a/r$. Left: mean radius $r_0$ in $a/m$ unit. Right: mass $M$ of the system in $m$ unit. \label{fig:coul}}
\end{center}
\end{figure}

We can expect a massive particle linked with a massless particle solely in a heavy-light meson, but then a confining potential is always present. So the system treated here with the sole Coulomb potential is quite academic. Nevertheless, equations are interesting to treat. With $V(r)=-a/r$, (\ref{AFM3fun}) becomes
\begin{equation}
\label{cp1}
Q + \frac{Q^2}{\sqrt{Q^2+m^2 r_0^2}} = a,
\end{equation}
which implies that $a-Q>0$. In this case, the solution is
\begin{equation}
\label{cpr0}
r_0 = \frac{Q}{m}\frac{\sqrt{a (2 Q-a)}}{a-Q}.
\end{equation}
The replacement of this value in (\ref{AFM1}) yields to
\begin{equation}
\label{cpM}
M = 2 m\sqrt{\frac{a}{2Q}\left( 1-\frac{a}{2Q} \right)}.
\end{equation}
This result was found in \cite{silv09} but without the discussion given in the following. As expected from scaling arguments, the ratio $M/m$ is only a function of $a$ and of the quantum numbers. This property is also valid for the genuine eigenvalue. This last result can be compared with the AFM solution of the symmetrical SSE (\ref{SSEsig}) with the Coulomb potential $-a/r$,
\begin{equation}
\label{cpMsym}
M^{(\sigma)} = \sigma m\sqrt{ 1-\left(\frac{a}{\sigma Q}\right)^2 }.
\end{equation}
This result can be found in \cite{hall02,hall03} or can be derived with the AFM  \cite{silv09}. As expected, in both cases, if $m$ is vanishing, the only energy scale disappears and the system cannot exist. So, the ultrarelativistic limit ($m \to 0$) is not relevant. But the asymmetrical case is more intriguing. Indeed, the rest energy is not properly defined in this case when the interaction disappears, since one of the particle is massless. From equations above, a solution $M$ exists only when the following condition is verified:
\begin{equation}
\label{cpcond}
\frac{a}{2} < Q < a.
\end{equation}
The right part of this inequality corresponds to the cancellation of the binding ($M\to m$ when $Q\to a$), and the left part corresponds to a collapse  ($M\to 0$ when $Q\to a/2$). This can be checked on fig.~\ref{fig:coul}. The mean radius $r_0 \to \infty$ when $Q\to a$, and $r_0 \to 0$ when $Q\to a/2$. A sufficiently strong interaction must exist to bind the system, but unphysical values of the mass could appear if it is too strong. With the Coulomb potential, the nonrelativistic limit is not defined by $m\to\infty$ but by $a\to 0$ (see (\ref{cpMsym})). So, (\ref{cpcond}) implies that this limit is irrelevant for the solution (\ref{cpM}).

The choice $p=-1$ ($P(r)=-1/r \propto V(r)$) implies that the AFM solution is an upper bound with $Q=Q_{-1}=n+l+1$. For instance, with $a=1.2$, only the ground state ($Q_{-1}=1$) exists. The AFM value (\ref{cpM}) gives $M/m = 0.9798$, while an accurate numerical calculation gives $M/m=0.8454$ \cite{brau98,sema01}. The upper bound is not very close to the exact value, but (\ref{cpM}) brings information about the (approximate) analytical structure of the solution.

\subsection{Linear potential}
\label{lp}

\begin{figure}[ht]
\begin{center}
\includegraphics*[width=0.45\textwidth]{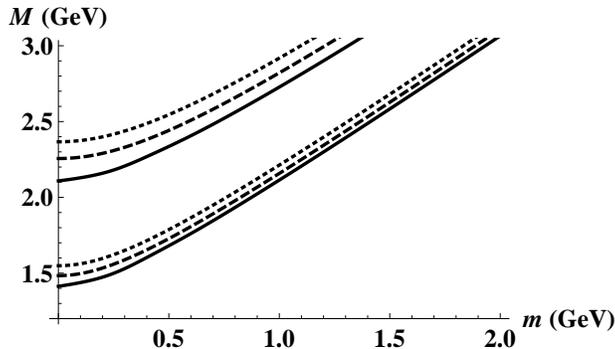}
\caption{Masses $M$ with quantum numbers ($l=0$, $n=0$) and ($l=0$, $n=1$) for a particle of mass $m$ interacting with a massless particle via a linear potential $b\,r$, with $b=0.2$~GeV$^2$. Solid line: accurate numerical solution \cite{brau98,sema01}; long-dashed line: AFM solution with $Q=Q_{1}$ ($P(x)=x$); short-dashed line: AFM solution with $Q=Q_{2}$ ($P(x)=x^2$). \label{fig:lin}} 
\end{center}
\end{figure}

A massive particle linked with a massless particle by only a linear potential can be considered as a reasonable approximation for a heavy-light meson. With a massless particle, the radius of the meson can be quite large, reducing the influence of the strong Coulomb potential. With $V(r)=b\,r$, (\ref{AFM3fun}) becomes
\begin{equation}
\label{cl1}
Q + \frac{Q^2}{\sqrt{Q^2+m^2 r_0^2}} = b\, r_0^2.
\end{equation}
The solution takes the following form
\begin{equation}
\label{clr0}
r_0 = \sqrt{\frac{Q}{b} -\frac{Q^2}{2 m^2}+ \frac{Q^{3/2}}{2 m^2} \sqrt{Q+\frac{4 m^2}{b}}}.
\end{equation}
It can be easily checked that the number under the square root is always positive. The replacement of this value in (\ref{AFM1}) yields to the AFM eigenvalue. We find more convenient to write it as a function of the AFM solution of the symmetrical SSE (\ref{SSEsig}) with $m=0$ and the linear potential $b\, r$ \cite{silv09}
\begin{equation}
\label{clMsym}
M^{(\sigma)}_0 = 2\sqrt{\sigma \, b\, Q}.
\end{equation}
After some algebra, we find
\begin{eqnarray}
\label{cpMAB}
&&A_\pm=\sqrt{\left( M^{(2)}_0 \right)^2 +32\, m^2}\pm M^{(2)}_0,\nonumber \\
&&B_\pm=\sqrt{M^{(2)}_0  A_\pm +16\, m^2}, \\
&&M=\frac{\sqrt{2}\, \left( M^{(2)}_0 \right)^2 A_- + 16\, m^2 \left(2 \sqrt{2}\, M^{(2)}_0 +B_+ \right)}{16\, m\,B_-}.\nonumber
\end{eqnarray}
It can be checked on fig.~\ref{fig:lin} that this quite complicated equation is an upper bound of the exact result for well chosen auxiliary potentials: $P(x)=x$ and $P(x)=x^2$.

For the linear potential, both ultrarelativistic and nonrelativistic limits are relevant. In the ultrarelativistic limit ($m \ll \sqrt{b})$, the dynamics is dominated by the motion of the two massless particles. Indeed, (\ref{cpMAB}) reduces to
\begin{equation}
\label{clMUR}
M \approx M^{(2)}_0 + \frac{2\, m^2}{M^{(2)}_0}.
\end{equation}
In the nonrelativistic limit ($m \gg \sqrt{b})$, the dynamics is dominated by the motion of the massless particle, the other one contributing quasi solely with its mass. As a matter of fact, the computation shows that (\ref{cpMAB}) reduces to ($M^{(2)}_0 = \sqrt{2}\, M^{(1)}_0$)
\begin{equation}
\label{clMNR}
M \approx m + M^{(1)}_0 + \frac{\left( M^{(1)}_0 \right)^2}{8\, m}.
\end{equation}
These two formulas can also be obtained with the AFM perturbation theory developed in \cite{sema11}. The ratios $M/M^{(2)}_0$ given by (\ref{clMUR}) and (\ref{clMNR}) coincide when $m/M^{(2)}_0 \approx 0.34$, and the relative error of both formulas with $M/M^{(2)}_0$ given by (\ref{cpMAB}) is then only 5.5\%. 

\section{Concluding remarks}
\label{conclu}

The spinless Salpeter equation is a semirelativistic equation mainly used to study two-body hadronic systems. Though a quite high precision can be reached for its numerical resolution, it is always interesting to have bounds on eigenvalues. A rapid estimation of the spectra is then possible, and sometimes analytical results can be obtained. As most of the works have been devoted to the study of the spinless Salpeter equation with two equal masses, we propose here an upper bound for systems with two different masses. 

We used the auxiliary field method whose concept, strongly connected with the envelope theory, is to replace a Hamiltonian $H$ for which analytical solutions are not known by another one $\tilde H$ which is solvable and which includes one or more auxiliary real parameters. Provided that the structure of the Hamiltonian $\tilde H$ is a nonrelativistic kinematics plus a power-law auxiliary potential, an eigenvalue computed by the auxiliary field method is simply the kinetic operator evaluated at the mean momentum per particle $p_0$ plus the potential energy computed at the mean radius $r_0$. The product $r_0\, p_0$ is equal to a global quantum number $Q$ depending on the auxiliary potential, and the value of $r_0$ is the solution of a transcendental equation. With a good choice of the auxiliary potential, the value of $Q$ is analytically known and the eigenvalue obtained is an upper bound of the genuine energy.

Numerical computations can be easily performed with this bound, but analytical closed formulas can also be obtained for some specific systems. Two such solutions are presented in the case where one particle is massless. The two potentials studied present some interest for the hadronic physics: The Coulomb potential and the linear confinement. Deep insights about the structure of the solutions can then be studied with the upper bounds.

\begin{acknowledgments}
CS thanks the F.R.S.-FNRS for financial support. 
\end{acknowledgments}

\end{document}